\documentstyle[12pt]{article}
\newcommand{\be}{\begin{equation}}
\newcommand{\ee}{\end{equation}}
\newcommand{\bq}{\begin{eqnarray}}
\newcommand{\eq}{\end{eqnarray}}

\begin{document}
\pagestyle{empty}

\begin{center}
{\Large \bf Theoretical Physics Institute}\\
%\vspace{0.2cm}
%University of Minnesota}
\end{center}
\begin{flushright}
{TPI-MINN-93/55-T}
\\
{UMN-TH-1227/93}\\
{Technion-PH-93-41}\\
{November 1993}\\
\end{flushright}
\vspace{.2cm}
\begin{center}

{\Large \bf Lifetimes of Charmed Hadrons Revisited. \\
Facts and Fancy
}
\end{center}

\vspace{0.2cm}
\begin{center}
{\em  Talk at the Third Workshop on the Tau-Charm Factory, \\
June 1993, Marbella, Spain }
\end{center}

\vspace{0.4 cm}
\begin{center}
{\large Boris Blok}

\vspace{0.2 cm}
{Department of Physics\\ Technion -- Israel Institute of
Technology\\
Haifa 32000, Israel
} \\

 \vspace{0.2 cm}
 and\\
 \vspace{0.2 cm}
{\large Mikhail Shifman}

\vspace{0.2 cm}
{ Theoretical Physics Institute,
 University of Minnesota \\ Minneapolis, MN 55455, USA}\\

 \end{center}
\vspace{0.4cm} \noindent

\begin{abstract}
The problem of  the hierarchy of lifetimes of charmed  hadrons is
reviewed.  The QCD-based theory of preasymptotic effects in
inclusive weak decays dating back to the beginning of the eighties is
now entering its mature phase.  Combining  recent and old results we
argue  that
the observed hierarchy reflects most intimate features of the
hadronic structure.
The problem of a wide spread of lifetimes of charmed hadrons is
addressed.
  We speculate on what is
to be expected from QCD to provide the observed pattern.  A number
of predictions is given for
the hierarchy of lifetimes in the family of the beautiful hadrons.
\end{abstract}

\newpage
\pagestyle{plain}
\setcounter{page}{1}

\section{Introduction}

\par From early days of QCD it is well-known that  the lifetimes of all
particles $H_Q$ containing a heavy quark $Q$ and light quarks
are the same and equal to that of a `free' heavy quark in the limit
$m_Q\rightarrow\infty$ ($m_Q$ is the mass of the quark $Q$). At
the same time the experimentally observed
total widths in the charmed family differ by as much as an order of
magnitude for $D^+$ and $\Xi^0_c$. This scatter clearly
shows that $m_Q\sim 1.4$
GeV is far below the asymptotic domain, and the questions arise
whether the
preasymptotic effects are understood in the present-day
QCD and what one should expect for the $b$ quark family.
These questions are addressed in the present talk. As we
will see later, the hierarchy of the lifetimes presents a junction
of different aspects of the theory of strong interactions. The two
main elements are (i) a systematic expansion of amplitudes in the
inverse mass of the heavy quark; (ii) estimates of matrix elements
of local operators (appearing in the expansion) over the hadronic
states $H_Q$. The first step, the construction of the $m_Q^{-1}$
expansion, takes into account the quark-gluon dynamics at
short distances and is clean in the sense that it is based on
fundamental
QCD and is, thus, model-independent (with a certain reservation, to
be discussed
below). The technology
 of $m_Q^{-1}$ expansions experienced a dramatic development in
recent years \cite{Bigi,Blok1}.
 The corresponding achievements are incorporated in our discussion.
As for the hadronic matrix elements, they reflect the hadron
structure at large distances. As usually, here the progress was
limited, if at all, and our purpose  is to outline some possible
approaches and  reveal the impact of this or that
specific mechanism on the lifetime hierarchy.

The  `scientific' approaches to the issue of the lifetime
hierarchy date back to the beginning of the eighties.
It was pointed out  that the total widths are related,
by means of unitarity, to the imaginary part of certain forward
scattering amplitudes \cite{Khoze}. This simple
and rather obvious
observation paved
the way to a generalization of the Wilson operator product expansion
(OPE) allowing one to represent all amplitudes of interest
in the form of a series in $m_Q^{-1}$
\cite{Voloshin1,Voloshin2,Tadic}, an
expansion essentially equivalent to what is called now the Heavy
Quark Effective
    Theory (HQET)
\cite{Hill,Georgi1} (for  reviews see e.g. ref. \cite{Georgi2}).
Shortly after
it was discovered that the terms behaving like powers of $m_Q^{-1}$
should be
supplemented by hybrid logarithms \cite{Voloshin3}. These
logarithms are
a routine part of the present-day HQET.
The analysis of preasymptotic effects in charm
and beauty combining both power series in $m_Q^{-1}$ and the
hybrid
logarithms has been first carried out in
\cite{Voloshin2} in
the mid-eighties.
At those days the data on baryons were scarce even in the
$c$ quark family, and next to nothing was known about the
lifetimes in the $b$ family. It is not surprising, therefore,
that the emphasis in these works was on the operators which
differentiate
between the $D^+$ and $D^0$. These operators necessarily involve
the spectator light quark and have dimension 6 or higher.
A set of dimension 6 operators was analyzed in
\cite{Voloshin2,Tadic} \footnote{Ref. \cite{Tadic}
ignores the existence of the hybrid logarithms.} with definite
predictions concerning the hierarchy of the lifetimes. Now, ten years
later, it became clear that the analysis has to be updated for many
reasons.

First, the numerical values of some crucial parameters turned out to
be higher
than previously thought. As a result, a typical scale of
the preasymptotic `corrections' becomes larger to the extent that the
`corrections' overshadow the leading term in the charmed
family (so that, strictly speaking, they can not be called
corrections anymore). Even the beauty family is still not quite
in the asymptotic domain.

The enhanced role of the preasymptotic effects leads, in
turn, to the necessity of including other operators, not fully
considered
in this problem previously. There are two operators
of dimension 5 and one new operator of dimension 6.
 Understanding  the
ratio of the baryon-to-meson lifetimes is impossible without
proper consideration of these operators.

As we will see below, for $m_c\sim 1.4$ GeV the convergence of
the expansion is unexpectedly {\em so bad} that even operators of
dimension 7 and
higher must play an important role. Not much can be said about
them at the moment, which means, of course, that all results
referring to the $c$ quark family are doomed to be valid at best
qualitatively, if at all. We have gloomy suspicions, to be
shared with everybody, that the QCD-based `inclusive' approach
will never become fully quantitative in the charmed family. In the
$b$ quark family the
situation seems to be more favorable. Here the
the expansion parameter $m_b^{-1} $ is sufficiently small to
guarantee a reasonable convergence of the $m_Q^{-1}$ series, and
limiting   to operators of dimension 5 and 6
is quite reasonable.

Next, one can use symmetry arguments and models
to estimate the hadronic matrix elements of these operators,
establishing
in this way a hierarchy of lifetimes.

This stage, evaluation of the matrix elements, is the most vulnerable
point of the existing theory, its Achilles' heel. One must frankly admit
that at least some of the relevant matrix elements are not controlled
theoretically today.  Hence,  one is forced to move onto  swampy soil
of models. We invoke a  primitive quark model for orientation.  What
is the outcome? Honestly, it should be admitted that the charmed
family is too light for serious theorists building $1/m_Q$ expansions.
With some imagination, however, we still dare to say that the pattern
of the charm lifetimes is understandable.  And further efforts, both
experimental and theoretical, are worthwhile --
the observed hierarchy (once it is actually established)
would imply certain relations between
the hadronic matrix elements of the operators involved, a unique
information, inaccessible in any other way.

Using the fact that this talk is not subject to the scrutiny of
refereeing we allow ourselves to indulge in a fantasy treating the
$1/m_c$ expansion and the primitive quark model mentioned  in a
rather speculative way. An element of wishful thinking is quite
obvious in this talk, and we bring our apologies for this
fact.  Our only justification is a hope that the corresponding
speculations will catalyze a more  `scientific' analysis in future.

We  start with a necessary theoretical background and a brief
historic
review.  Then we proceed to the picture of the $c$ quark
family.
Finally, a few remarks concerning our expectations in the $b$ quark
family are given.

\section{Theoretical Excursion}

\subsection{ General picture}

In prehistoric times (the seventies) the inclusive heavy flavor decays
were treated in  the most straightforward way -- the width of the
$Q$ containing hadron $H_Q$ was assumed to be equal to the
probability of the free $Q$ quark decay.  An immediate consequence
is  the equality of the widths of all hadrons belonging to the $Q$
family.

Although in the limit $m_Q\rightarrow\infty$ this prescription is
certainly correct  a single glance on what we actually have in the
charmed family (see Table 1)
shows how far the existing quarks are from the asymptotic domain
and how important preasymptotic corrections are. Even leaving aside
$\Xi^0_c$ where the data are rather poor and the error bars are
large we can say that  the lifetimes span the interval from roughly
$10\times 10^{-13}$ sec for $D^+$ (the most long-living charmed
particle) to $ 2\times 10^{-13}$ for
$\Lambda_c$ (see ref. \cite{Besson} for the recent analysis of the
experimental data).
  There is a drastic imparity between $D^+$ and $D^0$ and
between mesons and baryons.  Do we understand this hierarchy
theoretically?

The idea lying at the base of the modern theory of the preasymptotic
effects is simple and can be compared to a well-known text-book
problem. We mean the problem of the nuclear $\beta$ decay.
If the energy released in the $\beta$ electron is much larger than a
typical binding energy of both electrons in the shell and nucleons in
the nucleus (the latter is unrealistic, of course) one can forget about
the electrons in the shell and other bound state effects
in calculating the decay probability. In the high precision
calculations, however, the bound state effects
should be accounted for.  In particular, the soft electrons in the shell
do interfere with the $\beta$
electron (the Pauli exclusion principle), and this interaction generates
corrections proportional to
inverse powers of the energy release.

\subsection{The inverse quark mass expansion}

Likewise, in the heavy quark decay all participating degrees of
freedom can be either hard ( the corresponding energy is of order
$m_Q$) or soft.  In the leading approximation  we ignore the soft
degrees of freedom and treat the hard ones as free -- a good old free
quark decay (Fig. 1). The inclusion of the soft degrees of freedom in  analysis
generates  power non-perturbative corrections. An obvious example
of the soft degree of freedom is the spectator light quark.  This is not
the end of the story, of course. Soft gluons in the initial light cloud
surrounding $Q$
or those emitted by the final light quarks is another example (see
Fig. 2).

The formal basis allowing one to systematically catalog the
preasymptotic non-perturbative corrections is provided by Wilsonian
operator product expansion (OPE).
More exactly, what is readily calculable are the so called
condensate non-perturbative effects.

Although the word OPE might sound frightening to non-experts,
conceptually this is a very simple formalism, just a procedure of a
systematic separation   of large and short
distance contributions  in QCD (hard and soft degrees of freedom). We
further
assume
that the short-distance contribution residing in the coefficient
functions
can be found perturbatively, while all non-perturbative effects are
attributed to matrix elements of the operators appearing in the
expansion.
This procedure is sometimes called the practical version of OPE.  We
will refer to it as to the `standard'. It is clear that this standard
prescription is by no means exact
 since in principle there are non-perturbative contributions
even at short distances modifying the coefficient functions.
(They will be referred to as hard or non-condensate
non-perturbative terms).  The latter  are
 poorly controllable theoretically at present, and we will just assume
that they can be disregarded for the time being.  It seems quite
probable, though, that they are non-negligible in the charmed family.
Anyhow, we will disregard them.  The `hard'
non-perturbative terms should be (and probably will be) targeted in
future analyses
of the preasymptotic effects in the charm decays.

Constructing the operator product expansion  we
obtain the inclusive decay probabilities as an expansion in
 $m_Q^{-1}$. Practically all
terms up to $m_Q^{-3}$ are known at present. We will discuss them
and then speculate on higher order terms.

The statement above is an oversimplification. Apart from the power
dependence on $m_Q^{-1}$  one  encounters logarithmic
dependencies on this parameter -- the so called
 hybrid anomalous dimensions \cite{Voloshin3} (matching logarithms
of HQET). For pedagogical purposes let  us forget for a while about
the hybrid logs. We will comment on them later.

We start  from
the effective weak lagrangian describing non-leptonic decays
\begin{equation}
{\cal L}(\mu) = \frac{G_F}{\sqrt{2}}V_{sc}V_{ud}(c_1 {\cal O}_1 +
c_2 {\cal O}_2)
\label{9}
\end{equation}
where $c_{1,2}$ are Wilson coefficients accounting for the loop
momenta
from $\mu$ up to $M_W$ and known in perturbation theory
 while
${\cal O}_{1,2}$ are operators,
\begin{equation}
{\cal O}_1 = (\bar s\Gamma_\mu c)(\bar u\Gamma_\mu d) , \,\,\,
{\cal O}_2 = (\bar{s}_i\Gamma_\mu  c^j)(\bar{u}_j\Gamma_\mu
d^i).
\label{10}
\end{equation}
Here $\Gamma_\mu = \gamma_\mu (1+\gamma_5)$.  Eqs. (\ref{9}),
(\ref{10}) present
the lagrangian relevant to the $c\rightarrow s\bar d u$ transition.
The penguin
graphs showing up at the 1\%  level are  omitted.
The coefficients $c_{1,2}$ are known in the leading log and next-to-
leading approximations
\cite{Altarelli,Gaillard};
\begin{equation}
c_1 =\frac{1}{2}(C_++C_-),\,\,\, c_2 =\frac{1}{2}(C_+-C_-).
\label{11}
\end{equation}
In the leading approximation
\begin{equation}
 C_{\pm} = \left[
\frac{\alpha_s (\mu )}{\alpha_s (M_W)}\right]^{d_{\pm}}.
\label{7}
\end{equation}

The numerical values of the coefficients depend on the choice of
parameters.  The physical quantities, like the non-leptonic widths,
can not depend on the arbitrary scale $\mu$.   The $\mu$
dependence explicit in the weak lagrangian (\ref{9}) is compensated
in  the process of calculation of $\Gamma$'s.  In the leading log
approximation we limit ourselves to one can set $\mu = m_c$
from the very beginning. As for $\Lambda_{\rm QCD}$ the current
fashion bends towards a rather high value, $\Lambda_{\rm QCD}
\sim 300$ MeV
\cite{Altarelli2}. Our gut feeling is that the genuine  value of
$\Lambda_{\rm QCD}$ is actually smaller, but this is no time to
indulge in the discussion of this issue. Accepting this estimate we get
\begin{equation}
C_+\sim 0.74, \,\,\,  C_-\sim 1.8.
\label{eq:wilson}
\end{equation}

Now, we construct the transition {\em operator}
$\hat T(c\rightarrow X\rightarrow c)$,
\begin{equation}
\hat T = i\int d^4x T\{ {\cal L}(x) {\cal L}(0)\} =
\sum_i C_i {\cal O}_i
\label{eq:expansion}
\end{equation}
 describing a diagonal amplitude
with the heavy quark $c$ in the initial and final state (with identical
momenta). The transition operator $\hat T$ is built by means of OPE
as
an expansion in local operators ${\cal O}_i$.
The lowest-dimension operator in
$T(c\rightarrow X\rightarrow c)$ is $\bar c c$, and the complete
perturbative prediction corresponds to the perturbative calculation
of the
coefficient of this operator.  In calculating the coefficient of $\bar c
c$ we treat the light $s,d,u$ quarks in $X$ as hard and neglect the
soft modes. Say, we ignore the fact that in a part of the phase space
the $u$ quark line is soft and can not be treated perturbatively.
Likewise, we ignore interaction with the soft gluons.

Our task -- the task of the theory of preasymptotic corrections --
is the analysis of the influence
of the soft modes in the quark and gluon fields manifesting
themselves
as a series
of high-dimension operators in $\hat T$.

Once the expansion (\ref{eq:expansion}) is built we average $\hat T$
over the
hadronic state of
interest,
charmed  mesons and baryons in the case at hand. At this stage the
non-perturbative
large distance dynamics enters through the matrix elements of the
operators of dimension  5 and higher. (There are
no operators of dimension 4).

Finally, the imaginary part of $<H_c|\hat T|H_c>$ presents the $H_c$
width sought for,
\begin{equation}
\Gamma (H_c)=\frac{1}{M_{H_c}}
<H_c\vert {\rm Im} \hat T(c\rightarrow X  \rightarrow c)\vert H_c>.
\label{eq:width}
\end{equation}

For experts we hasten to add a  remark concerning peculiarities of
the kinematics
of the amplitude under consideration. It is essentially Minkowskian.
Still the operator product expansion originally formulated for
deep Euclidean kinematics can be used. It is important that we take
the full imaginary part corresponding to all possible cuts of the
diagrams (totally inclusive final state $X$). Omitting some of the cuts
would lead to infrared unstable results \cite{BU}.
Since the decay rates we calculate  present integrated quantities
over the interval of energies up to $m_c$ the corresponding
prediction for $\Gamma$'s,
through analyticity, is related to the calculation of the OPE
coefficients
with the characteristic off-shellness $\sim m_c$.  For comparison let
us
mention a similar analysis of the $\tau$ decays \cite{Pich}.

\subsection{Catalog of relevant operators}

The leading operator in the expansion of $\hat T$  is
\begin{equation}
{\cal O}_0=\bar c c.
\label{14}
\end{equation}
Keeping only $\bar c c$ is equivalent to perturbative calculation of
the
coefficients $c_{1,2}$ followed by a perturbative
calculation of the decay width $\Gamma$ through the
relation
\begin{equation}
\Gamma^{pert} = \frac{1}{m_c} {\rm Im} <c|\hat T|c>,
\label{15}
\end{equation}
where  taking a `mathematical' matrix element over the $c$ quark
singles
out the operator $\bar c c$.

It is worth noting that the contribution of ${\cal O}_0$ in the
physical width $\Gamma$ (not to be confused with
$\Gamma^{pert}$!)
is given by
$$
\Gamma =2{\rm Im}C_0\frac{1}{2M_{H_c}}<H_c|{\cal O}_0|H_c>.
$$
The matrix element $(1/2M_{H_c})<H_c|{\cal O}_0|H_c>$ is 1 plus ${\cal
O}(m_c^{-2})$
corrections (see eq. (12) below).

The next operator in our catalog  is
\begin{equation}
{\cal O}_G=\frac{i}{2}\bar c \sigma_{\mu\nu}
G_{\mu\nu} c\rightarrow - \bar c \vec\sigma\vec B c ,
\label{16}
\end{equation}
with  dimension 5. This operator as well as ${\cal O}_\pi$, see below,
generates a difference between mesons on one hand and baryons, on
the other.

Let us make a side remark  concerning another dimension 5 spin 0
operator,
\begin{equation}
{\cal O}_\pi =\bar c [D^2 -(vD)^2] c,
\label{21}
\end{equation}
where $v$ is the four-velocity of the hadron $H_c$.
This operator does not
 appear explicitly in the calculation of the
total widths due to the fact that ${\cal O}_\pi$ is not Lorentz scalar.
On the other, the relation \cite{Bigi,BSUV}
\begin{equation}
\bar c c =\bar c\gamma_\mu v_\mu c - \frac{1}{2m_{c}^{2}}{\cal
O}_\pi +
\frac{1}{2m_{c}^{2}}{\cal O}_G +
\frac{1}{4m_c^3}g^2 \bar c \gamma_0 t^a c\sum_q \bar q \gamma_0
t^a q+
O (1/m_{c}^{4})
\end{equation}
used  in the calculation of the matrix element $<H_c|\bar c c|H_c>$
contains ${\cal O}_\pi$. (Here $t^a$ are the color generators and $q$
is a generic notation for light quarks).

 We pass now to  the ${\cal O}(m_c^{-3})$ terms due to
operators of dimension 6. On general grounds one can limit the
number
of these operators to a few following structures \cite{BSUV}.

(i) {\em Four-quark operators}

These have the generic form
$$
{\cal O}_{4q}=(\bar c \Gamma q)(\bar q\Gamma c)
$$
where
$q$ is a light quark field and $\Gamma$ presents, in the case at
hand, a
combination of the Lorentz and color matrices.  The coefficients of
the four-quark operators have been first calculated in refs.
\cite{Voloshin1,Voloshin2,Tadic}.
 They are determined by the one-loop graphs of the
type presented on Fig. 3.

The four-fermion operators are responsible for the splittings
between $D^+$ and $D^0$ and the splittings between different
charmed baryons.

(ii) {\em Dimension 6 quark-gluon  operators}

Apart from the heavy quark fields $c$ and $\bar c$ relevant
operators
contain the gluon field strength tensor $G_{\mu\nu}$
(they may also include the
covariant derivatives). These operators are generated by two-loop
graphs
(fig. 2) and, hence, their coefficients are numerically strongly
suppressed
compared to the four-quark case above.

By using equations of motion it is not difficult to show
\cite{BSUV} that there are only two options for
the spin-zero quark-gluon operators of dimension 6, namely
$$
\bar c (D_\mu G_{\mu\nu})\Gamma_\nu c
$$
and
\begin{equation}
{\cal O}_E = \bar c\sigma_{\mu\nu}G_{\mu\rho}\gamma_\rho
iD_\nu c
\rightarrow \bar c\vec\sigma \, \vec E \times i \vec D c ,
\label{25}
\end{equation}
where $\vec E$ is the chromoelectric field. All other operators
of dimension 6 with
the gluon field are  reducible to the above.
The operator $\bar c (D_\mu G_{\mu\nu})\Gamma_\nu c$ is actually
a four-quark operator since
$$
D_\mu G_{\mu\nu} =  - g^2\sum \bar q \gamma_\nu T^a q .
$$
Its coefficient contains extra $\alpha_s/\pi$, however, compared
to the four-quark operators coming from the one-loop graphs,
and the corresponding contribution can  be neglected.

As for ${\cal O}_E$  its contribution turns out to be  small
numerically compared to that of ${\cal O}_G$, so that it plays no role
in what follows. We mention it for completeness and will essentially
forget about  ${\cal O}_E$ from now on.

For pedagogical purposes it is instructive to write down  the
decomposition of the operators above in terms of the HQET field,
\begin{equation}
h_c(x)=e^{-im_cvx}\frac{(I+\hat v)}{2} c(x),
\label{eq:heavy}
\end{equation}
although, certainly,
this decomposition  adds nothing in practical terms.
In eq. (\ref{eq:heavy}) $c$ is the normal Dirac bispinor, and $v$ is
the
four-velocity of the heavy hadron; $\hat v\equiv
\gamma^{\alpha}v_\alpha$.

Then, say, for $\bar c c$ we  essentially repeat eq. (12),
\begin{equation}
{\cal O}_0 \rightarrow \bar h_c\gamma_\mu v_\mu h_c -
\frac{1}{2m_{c}^{2}}\bar h_c{\vec\pi}^2 h_c -
\frac{1}{2m_{c}^{2}}\bar h_c\vec\sigma\vec B h_c + ...
\label{eq:t}
\end{equation}
and
\begin{equation}
{\cal O}_G\rightarrow -\bar h_c\vec\sigma\vec B h_c
+...
\label{eq:thirdsource}
\end{equation}
It is possible to prove that all dimension 6 operators  appearing  in
eqs.
(\ref{eq:t}), (\ref{eq:thirdsource})
 can be reduced to four-quark operators
suppressed by $\alpha_s$. They  can be consistently discarded.

\subsection{Coefficients}

The operators listed above appear in $\hat T$ each with its own
coefficient. Practically all of them have been  calculated previously --
for ${\cal O}_0$ in the prehistoric times. The coefficients of the
operators of dimension 5 were found
in the recent works \cite{Bigi,Blok1} and for the spectator-sensitive
operators of dimension 6 in refs. \cite{Voloshin1,Voloshin2}.
We supplement this set by a new spectator-blind operator of
dimension 6,
and consider a new source of $1/m_Q$ corrections -- the expansion
of matrix
elements of operators of dimensions 3 and 5 in terms of $1/m_Q$.

The prehistoric coefficient $C_0$ is obviously equal to
\begin{equation}
{\rm Im} C_0 =\frac{1}{2} (3\Gamma_0 )\eta
\end{equation}
where
\begin{equation}
\Gamma_0\equiv \frac{G_F^2m_c^5|V_{sc}|^2}{192\pi^3}
\label{3}
\end{equation}
is a very convenient unit for all widths to be discussed here. The
factor $\eta$ reflects the hard gluon exchanges hidden in  $C_{\pm}$,
\begin{equation}
\eta = \frac{C_-^2+2C_+^2}{3} \approx 1.5.
\label{7*}
\end{equation}
If the hard-gluon exchanges are accounted for in the next-to-leading
order
$$
\eta\rightarrow\eta J
$$
where the explicit expression for $J$ is given in ref. \cite{Petrarca},
see also \cite{Gaillard}.
Since the picture of the lifetime hierarchy we are aimed at is
qualitative at best we limit ourselves to the leading approximation.

The strange quark mass is neglected,
here and below, systematically. Since we deal exclusively with the
current quarks (there is no such notion as the constituent quark in
the QCD-based approach),
$$
m_s^2/m_c^2 \sim 10^{-2}
$$
and the approximation $m_u=m_d=m_s=0$ in calculating
the {\em inclusive} widths seems to be not bad. One should be alert,
however -- possible deviations from this limit might be larger than
expected, and an estimate of the effects due to non-vanishing
$M_K^2$ is definitely in the short list of questions for further
investigation.

Now we are approaching a more interesting and less trivial stage of
the analysis -- we pass to soft modes.  (A reminder:  we discard
non-condensate non-perturbative effects. This is not because they
are necessarily small in the charm decays but because we do not
know how to treat them. Thus, here is another item from the short
list -- hopefully, a courageous theorist can be found for its
investigation.)
The first effect to show up is
that due to the chromomagnetic operator $\bar c \vec\sigma\vec Bc$
describing the correlation of the heavy quark spin in $H_c$ with the
chromomagnetic field in the light cloud surrounding $c$. Its
dimension is 5, hence the corresponding contribution in $\Gamma$
is ${\cal O}(m_c^{-2})$.

The chromomagnetic field $\vec B$ appears because both the initial
quark $c$ and the light fast quarks produced in the decay are
coupled to it -- they do not live in the empty space but, rather, in a
medium, the `brown muck' around the heavy quark (we prefer
to call this medium `light cloud').  There are two sources of ${\cal
O}_G$: the direct coupling to $\bar d$ and the coupling to $c$.  In the
latter case  ${\cal O}_G$ appears through the equations of motion
for
the $c$ field.  We will not dwell on the issue referring to the original
publications \cite{Bigi,Blok1,DPF}
 for further details.  To the leading order in
$\alpha_s$ the coefficients $C_G$ and $C_\pi$ are
\begin{equation}
C_G = -\Gamma_0 \frac{1}{m_c^2}(8c_+^2- 2c_-^2),
\end{equation}
\begin{equation}
C_\pi =0.
\end{equation}
${\cal O}_\pi $ appears only when we take  the matrix element of
${\cal O}_0$, along with an additional piece with ${\cal O}_G$, see
eq. (12).

Although both operators, ${\cal O}_G$ and ${\cal O}_\pi$,
are of dimension 5 their matrix elements
over $H_c$ may contain also corrections ${\cal O}(m_c^{-1})$
corresponding to ${\cal O}(m_c^{-3})$ terms in $\Gamma$'s
which turn out to be insignificant numerically.

We now move to the operators of dimension 6.
As we argued above  there are only two classes of operators
that we have to consider here. All other operators can be reduced
(through
equations of motion) to color blind four-quark operators, whose
Wilson  coefficients
are additionally suppressed by $\alpha_s$. Since considering
$\alpha_s$ corrections to the Wilson coefficients is beyond our
accuracy now
we shall neglect them. We are  left  with two
classes of operators.
First, and most important,  are spectator-sensitive four-quark
operators,
 whose
analysis was carried out in refs. \cite{Voloshin1,Voloshin2,Tadic}.
Their coefficients are numerically enhanced since they are generated
by one-loop graphs.
These operators lead to the lifetime differences between different
mesons
and different baryons. There are three distinct physical mechanisms
summarized by the four-quark operators: annihilation, interference
and quark-quark scattering in the baryons. Technically the
distinction manifests itself in the fact that each of the three light
quark lines, $u$, $d$ or $s$ can be soft.  Pictorially the softness of the
line is depicted as follows: instead of drawing a solid line we just cut
the soft line to remind that no perturbative expression can be used
for this line.  Three relevant graphs are given on fig. 3 a,b,c.
 The full contribution of the 4-quark operators to the transition
operator is
\begin{equation}
\hat T_{4q}={\cal L}_d+{\cal L}_u+{\cal L}_s .
\label{eq:4quark}
\end{equation}
Here
\begin{equation}
{\cal L}_d= 48\pi^2\Gamma_0\frac{1}{m_c^3}(K_1(\bar c
\Gamma_\mu c)
(\bar d \Gamma_\mu d)+K_2(\bar c \Gamma_\mu
d))
(\bar d \Gamma_\mu c)) ,
\label{eq:firstterm}
\end{equation}
\begin{eqnarray}
\begin{array}{cl}
{\cal L}_u&=48\pi^2\Gamma_0\frac{1}{m_c^3}(K_3(\bar c
\Gamma_\mu c
-\frac{2}{3}\bar c \gamma_\mu\gamma_5c)(\bar u\Gamma_\mu u
)\\
&+K_4(\bar c_i\Gamma_\mu c_k-
\frac{2}{3}
\bar c_i\gamma_\mu\gamma_5c_k)(\bar u_k\Gamma_\mu u^i)) ,
\label{eq:secondterm}
\end{array}
\end{eqnarray}
\begin{eqnarray}
\begin{array}{cl}
{\cal L}_s&=48\pi^2\Gamma_0\frac{1}{m_c^3}(K_5(\bar c
\Gamma_\mu c
-\frac{2}{3}\bar c \gamma_\mu\gamma_5c)(\bar s\Gamma_\mu s
)\\
&+K_6(\bar c_i\Gamma_\mu c_k-
\frac{2}{3}
\bar c_i\gamma_\mu\gamma_5c_k)(\bar s_k\Gamma_\mu s^i)) ,
\label{eq:thirdterm}
\end{array}
\end{eqnarray}
where
\begin{eqnarray}
\begin{array}{cl}
 K_1&= (c_+^2+c_-^2)/2 ,\\
 K_2&=(c_+^2-c_-^2)/2 ,\\
 K_3&=-(c_++c_-)^2/8 ,\\
 K_4&= -(5c_+^2+c_-^2-6c_+c_-)/8 ,\\
 K_5&= -(c_+-c_-)^2/8 ,\\
 K_6&= - (5c_+^2+c_-^2+6c_+c_-)/8 .\label{eq:coefficients}
\end{array}
\end{eqnarray}

The term ${\cal L}_d$ describes
in the case of the $D^+$ meson
 the destructive interference between
the $\bar d$
quark produced in the charmed quark decay and the spectator $\bar
d$ quark. The same operator in  the case of  $\Lambda_c$
describes
the $cd\rightarrow us$ scattering. The operators ${\cal L}_u$ and
 ${\cal L}_s$ describe
the interference between the spectator quarks in baryons and the
$u$ or $s$ quarks produced in the decay (this interference is
destructive
for the
case of the
 $u$ quark and constructive for the case of the $s$ quark).

\subsection{Hybrid anomalous dimensions}

Above the coefficients were presented for the operators
normalized
at $m_c$. Generally speaking,  these operators have the hybrid
anomalous
dimensions \cite{Voloshin3,Voloshin2}
 reflecting the effect of the virtual momenta
from $m_c$
to a typical hadronic scale $\mu$.
If we know the hadronic matrix elements of the operators
normalized at $m_c$ we do not need to include the hybrid logs
explicitly, they are already there. As we will see shortly  this is the
case with the operator ${\cal O}_G(\mu =m_c)$ whose matrix
element over mesons is expressible in terms of the measured mass
differences while the matrix elements over $\Lambda_c$, and
$\Xi_c$ vanish.  The
operators $\bar c c $ and $\bar c \vec D^2 c$ have zero hybrid
 anomalous dimensions.

As for the four-fermion operators, however, we have no independent
information on their matrix elements.  Thus we have to use models
-- vacuum saturation in mesons and quark models in baryons.
The question arises as to what
choice of the normalization point would ensure  the best possible
validity of these models.

 As far as the strong
interactions are concerned, $m_c$ is a completely foreign
parameter.
Therefore, it is absolutely natural to think that the models  work
best
for a low normalization point, $\mu\sim$ a few hundred MeV.  Then
one should evolve the
four-fermion
operators down to $\mu$, include explicitly the hybrid anomalous
dimensions and only then apply the models. The effect of the hybrid
logarithms
is known in the leading-log approximation \cite{Voloshin2}.

Taking account of the hybrid logs changes the coefficients $K_i$
introduced in eq. (\ref{eq:coefficients}) as follows,
\begin{equation}
K_i \rightarrow K_i\eta _i
\end{equation}
where
\begin{eqnarray}
\begin{array}{cl}
K_1\eta_1 &= (c_+^2+c_-^2+\frac{1}{3}
(1-\sqrt{\kappa})(c_+^2-c_-^2))/2 ,\\
K_2\eta_2&=\sqrt{\kappa}(c_+^2-c_-^2)/2 ,\\
K_3\eta_3&=-[(c_++c_-)^2+\frac{1}{3}(1-\sqrt{\kappa})
(5c_+^2+c_-^2-6c_+c_-)]/8 ,\\
K_4 \eta_4&=- \sqrt{\kappa}(5c_+^2+c_-^2-6c_+c_-)/8 ,\\
K_5\eta_5&=-[(c_+-c_-)^2+\frac{1}{3}(1-\sqrt{\kappa})
(5c_+^2+c_-^2+6c_+c_-)]/8 ,\\
K_6 \eta_6 &=-\sqrt{\kappa}(5c_+^2+c_-^2+6c_+c_-)/8 .
\label{eq:factors}\\
\end{array}
\end{eqnarray}
Here $$\kappa=\alpha_s(\mu)/\alpha_s(m_c) .$$
The evolution from $m_c$ down to $\mu$ also adds new structures,
for instance,
\begin{equation}
\Delta {\cal L}_d =16\pi^2\frac{1}{m^3_c}\Gamma_0
(c_+^2-c_-
^2)\kappa^{1/2}(\kappa^{-2/9}-1)
(\bar c \Gamma_\mu t^ac)j^a_\mu  ,
\label{eq:s1}
\end{equation}
\begin{equation}
\Delta {\cal L}_s =-8\pi^2\frac{\Gamma_0}{m^3_c}
\kappa^{1/2}(\kappa^{-2/9}-1)(5c_+^2+c_-^2)(\bar c
\Gamma_\mu
t^a
c-\frac{2}{3}\bar c\gamma_\mu\gamma_5t^ac)j^a_\mu .
\label{eq:s2}
\end{equation}
Here
$$j^a_\mu=\bar u\gamma_\mu t^au+\bar d\gamma_\mu t^ad+\bar
s\gamma_\mu t
^a s .$$
The latter structures
 are  similar to the penguin contributions in the
weak charm decays. Since we neglected the penguins, it seems logical
to discard the structures of the type (\ref{eq:s1}), (\ref{eq:s2}) as
well.
We refer the reader to ref. \cite{Voloshin2} for a detailed
discussion of the hybrid logarithms in the transition operator.
\par Apart from the logarithmic renormalization of the currents and
operators discussed above, hybrid logarithms also appear in the
matrix
elements of the 4-quark operator when we use the factorization
approximation to calculate them. We shall return to this point in the
next
section.
\par Finally let us note that although the dimension 5
chromomagnetic
operator ${\cal O}_G$ has non-zero hybrid anomalous dimension,
we do not have to take into account the hybrid renormalization when
we
analyses its contribution. The reason is that the  matrix element of
${\cal O}_G$ normalized at $m_c$
will be expressed in the next section through physical quantities
-- the masses of the charmed hadrons .

\section{The hierarchy problem}
\par The most striking feature of the experimental data on the
lifetime
hierarchy obvious from the first glance  at Table 1 is its wide
spread, especially for the case of the baryons.
 This directly contradicts our intuition. Indeed, the  known
source of the differences in lifetimes between different baryons are
the 4-quark operators. Although their coefficients are  enhanced by a
factor
$4\pi^2$  their matrix elements are suppressed by a factor
$f^2_D/M^2_D\sim
0.01$. (Here $f_D$ is the  value of the  decay constant
of the D meson). Hence, one could  expect the lifetime spread in the
hierarchy to be
at most of
order one. Certainly one could not forsee that the lifetimes will span
an  order of magnitude, as
 experimental data show. The lifetimes differing by 50 to 100\% was
the prediction of the
first works  on this subject \cite{Voloshin1,Voloshin2}. In addition to
this, the analysis \cite{Voloshin1,Voloshin2} had other problems. For
instance,
it was obtained that the lifetimes of  $\Lambda_c$ and $\Xi^+_c$ are the
same
irrespective of the absolute value of $<{\cal O}_{4q}>$. This result
seems to contradict
the experimental data.
Some progress in understanding of the latter  problem was achieved
in ref.
\cite{Tadic}, where it was noticed that the relative weights of
 different 4-quark operators are very sensitive to the Wilson
coefficients $c_+$ and $c_-$. The use of the  modern value of
these coefficients corresponding to
 $\Lambda_{\rm
QCD}\sim 250-300 $ MeV
substantially
 improves  the situation with   $\Lambda_c /\Xi^+_c$ compared to
the
earlier results of ref. \cite{Voloshin2}.
Still the problem of the wide spread  remains. Can we
explain it from  first principles?
 The remainder of this talk will be devoted to an  explanation
(or, more precisely to speculations) why such a hierarchy can be
expected
from QCD (or, more precisely, how  QCD must work so that the theory
will
be able to explain the experimental data).

\section{What Can Be Said about Mesonic and Baryonic Matrix
Elements}

Unfortunately, the issue of the matrix elements, the most
underdeveloped element of the whole procedure at present, can not
be deferred indefinitely. We have to address it, realizing, of course,
that now we are not going to enjoy the same degree of theoretical
control  we had previously.

For mesons the situation is not that bad, though, as  will be seen
below.
At the same time in baryons we have to use much more shaky
models.

For each operator in the OPE we have discussed  we consider  first
mesonic matrix elements and then, later on,
the matrix elements over the baryon states.

\subsection{Mesons}

The first operator of dimension 5 is the operator $\bar c\sigma G c$. The
matrix elements of this operator between the mesonic states were
discussed
in detail in ref. \cite{Blok2}.  Remarkably, this matrix element can be
determined in a model independent way, using the HQET
symmetries:
\begin{equation}
\frac{1}{4M_D}<D\vert \bar ci\sigma G c(\mu = m_c )\vert D>\equiv
\mu^2_G=\frac{3}{4}(M^2(D^*)-M^2
(D))\sim 0.4 {\rm GeV}^2.
    \label{eq:value}
\end{equation}

The next relevant matrix element is the operator of the kinetic
energy \begin{equation}
-\frac{1}{2M_D}<D\vert \bar c\vec D^2c\vert D>\equiv
\mu^2_\pi\sim 0.5 {\rm
GeV}^2 .
\label{eq:kinetic1}
\end{equation}
Here we used the QCD sum rule results \cite{Braun}.

We now move to the dimension 6 operators.
Just one remark to demonstrate that  ${\cal O}_E$ does not play a
role.
Indeed, assuming that the chromoelectric field is of the same order
as the
chromomagnetic field, we obtain that the matrix element of this
operator
is suppressed by a factor $\sim \mu_\pi/m_c\sim 0.4$ in
comparison with
the contribution of dimension 5 operators.
The corresponding Wilson coefficient is
$$c_E=4\Gamma_0/m^3_c $$
 The contribution of this dimension 6 operator is less than $ 10\%$ of
the
contribution of the dimension 5 operators and can be safely neglected.

Let us turn now to the four-quark operators.
The simplest approach allowing one to estimate the matrix elements
of these
operators is
to use factorization.  Consider first the
operator $\bar c
\Gamma_\mu d
\bar d \Gamma_\mu c$.
Note first that this matrix element contributes only to the $D^+$, but
not
to the $D^0$ meson.
 This matrix element can be estimated using factorization, or the
vacuum
saturation  method:
\begin{equation}
<D^+\vert \bar c \Gamma_\mu d
\bar d \Gamma_\mu c\vert D^+>=f_D^2M^2_D
\label{eq:fact}
\end{equation}
where $f_D$ is the $D$ meson decay constant,
\begin{equation}
<0\vert\bar c \Gamma_\mu d \vert D> =f_D ip_\mu .
\label{eq:decay}
\end{equation}
The operator $\bar c \Gamma_\mu d
\bar d \Gamma_\mu c$
and  $f_D$ are both normalized at one and the same point, $m_c$.
If we would like to proceed to $f_D$ normalized at $\mu$
the corresponding hybrid logarithm should be inserted.

We pause here to make a remark which may be very important
numerically. The question is which coupling $f_D$ is to be
substituted in eq. (\ref{eq:fact}).  The physical coupling $f_D$
contains in itself
corrections $m_c^{-1}$ that are known to be very significant.
Therefore, if we want to have a systematic $1/m_c$ expansion we
must use a static value of $f_D$, with the $1/m_c$ terms removed.
The
difference between the physical and the static value of $f_D$ clearly
manifests itself in our expansion for the widths as an  effect ${\cal
O}(m_c^{-4})$. Were the expansion well convergent whatever value is
substituted would not matter. For the actual $c$ quark this  matters
a lot,
since the physical value of $f_D$ is
$\sim 170$ MeV, while the static $F_D\sim 400$ MeV, i.e. $\sim$2.5
times
larger
than the physical one \cite{F,F1}. The difference between
$f_D^2$ and $F_D^2$ is humongous.
The static coupling constant is defined as follows
\begin{equation}
<0\vert \bar d \Gamma_\mu h\vert D>=iF_Dv^\mu\sqrt{M_D/2} .
\label{eq:static}
\end{equation}
We will use capital $F$ to emphasize the distinction between the
physical
and the static constants.
$F_D$ defined by eq. (\ref{eq:static})  depends on the mass
of the heavy quark as
\begin{equation}
F_D\sim \vert \psi (0)\vert^2/\sqrt{m_c}
\label{eq:fa}
\end{equation}
modulo hybrid logs.
Here $\psi(0)$ is the value of the heavy meson wave function at the
origin that
does not depend on the mass of the heavy quark. If  the
value
of matrix element (\ref{eq:fact})  calculated with $F_D$ is
substituted in our OPE  we see that the
corresponding
contribution is pure $1/m^3_c$ compared to  perturbation
theory.

On the other hand, the leptonic decay constant $f_D$ defined using
eq.
(\ref{eq:decay}) contains all orders of $1/m_c$. Indeed, the
relativistic
Dirac field $c(x)$  describing the $c$ quark in QCD is connected
with
the
the $c$ quark field in  HQET as
\begin{equation}
c(x)=e^{-im_cvx}\Sigma_k\frac{(i\hat D)^k}{(2m)^k}h_c .
\label{eq:HQET}
\end{equation}
Here $\hat D\equiv \gamma^\alpha D_\alpha$.
Consequently, the static and physical decay constant are connected as
\begin{equation}
f_D=F_D(1+C_1/m_c+...),
\label{eq:connection}
\end{equation}
$C_1$ is a constant  independent of the heavy quark
mass.  It can be (and was) determined
in the QCD sum rules \cite{F}  and on the
lattices \cite{F1}.  It turns out
that
the contribution of the term proportional to $C_1$  in eq.
(\ref{eq:connection}) is approximately 50$\%$ of the contribution of
the
leading term and has the  negative sign.

Only a systematic analysis of all dimension 7 operators can tell us
what value -- physical versus static -- is more relevant to the
problem. In the absence of such an analysis we will just speculate
without even pretending on seriousness.

Other four-quark matrix elements can be determined also using
factorization and the Fierz identities. In particular,
\begin{equation}
<D \vert\bar c \Gamma_\mu t^a d
\bar d \Gamma_\mu t^a c\vert D>\approx 0.
\label{eq:zero}
\end{equation}

For the product of two color-singlet brackets factorization becomes
exact in the limit of the infinite number of colors.

The natural question is how well factorization works in the real
world.
The corresponding estimate is completely analogous to the estimates
of the $\bar D-D$ mixing parameters where factorization works with
the accuracy of order
 $20\%$. This means that for mesons we can determine
matrix elements in a model independent way with a sufficiently high
accuracy.

\subsection{Baryons}

We now move to baryons.
Consider first the dimension 5 operators. We start from the operator
$\bar c\sigma G c$. The matrix elements of this operator
between the
members of the antitriplet of the charmed baryons are zero. Indeed,
this
matrix element is proportional to correlation between the spin of
the
heavy quark and that of the
light cloud. The latter spin in the baryons
-- members of the antitriplet is zero.

 An estimate of this matrix
element
for the baryon $\Omega_c$ belonging to the sextet of the charmed
baryons
(and so the
light cloud there has spin 1) can be carried out in
the same
way as for mesons. Recall that a part of the Hamiltonian of the heavy
quark
effective theory  responsible for the spin splittings between the
hadrons has the form
\begin{equation}
\Delta H_{\rm eff}=\frac{h_c\vec\sigma \vec B h_c}{2M} .
\label{eq:split}
\end{equation}
Since the chromomagnetic field is due to the light cloud, it is quite
obvious
that
\begin{equation}
\vec B=\frac{C}{2}\vec S_{\rm l.c.},
\label{eq:spin}
\end{equation}
where $C$ is a constant and 1/2 is introduced for convenience.
Then we easily obtain that
\begin{equation}
\Delta M^2(\Omega_c^{(3/2)})\equiv <\Omega_c^{(3/2)}\vert {\bar
h}_c \vec\sigma\vec B h_c\vert
\Omega_c^{(3/2)}>
=\frac{1}{2}C
\end{equation}
and
\begin{equation}
\Delta M^2(\Omega_c)
\equiv <\Omega_c\vert\bar h_c \vec\sigma\vec B h_c\vert
\Omega_c>
=-C .
\end{equation}
Here $\Omega_c^{(3/2)}$ is the corresponding member of the baryon
sextet
with
the spin equal to 3/2. As a result, the matrix
element of the
operator ${\cal O}_G$ is given by
\begin{equation}
\frac{1}{2M_{\Omega_c}}<\Omega_c|{\cal O}_G|\Omega_c>=-
 <\Omega_c\vert \bar h_c\vec \sigma\vec B h_c\vert \Omega_c>
=\frac{2}{3}(M^2(\Omega_c^{(3/2)})-M^2(\Omega_c)) .
\label{eq:T}
\end{equation}
The  mass difference $M(\Omega_c^{(3/2)})-M(\Omega_c)$
can be determined using nonrelativistic quark
model \cite{Gleshow} and is $\sim 50$ MeV.

Consider now the matrix element of the operator $
{\cal O}_\pi = -\bar c\vec D^2c$.
The
matrix element of this operator is nothing else than the average
momentum
square of light degrees of freedom.  In ref.
\cite{Manohar} it is argued  that the matrix elements of ${\cal O}_\pi
$
are essentially independent of  the particular  hadronic state  and are
approximately  the
same for
all $c$ containing hadrons. In the absence of better ideas we accept
this statement for orientation.
If so, one can use the estimate of this matrix element from
 ref. \cite{Braun}.

The most difficult task in the baryon sector refers to the matrix
elements of the four-quark operators. Unfortunately, in this
case
we do not have such a strong tool  as  factorization applicable  in the
case of
mesons. In fact, at present we do not know any model-independent
approaches permitting  one to determine or even to estimate the
matrix
elements over the baryons. In order to say at least
something
in this case we have to turn to different models that are not
based on first principles  and are related to QCD only remotely, if at
all.
Consequently, the estimates of the matrix elements will be beyond
theoretical control.

In the literature analysis of the relevant matrix elements over the
baryonic
states was carried out  only in the  nonrelativistic
quark
model (NQM) \footnote{In NQM non-relativistic limit is assumed for
both, $c$ quark and the light ones from the cloud.} and its close
relatives, such as the bag model.  The former will be discussed at
length. Numerically the bag model predictions turn out to be rather
close. As a matter of fact, essentially one feature of NQM is crucial --
the absence of the spin-spin terms in the  matrix elements of ${\cal
O}_{4q}$ over
 the triplet baryons (for brevity this clumsy combination of words is
substituted by a shorthand `spin-spin interactions'). The
overall normalization of the matrix elements does not matter much
since effectively it is adjusted in an {\em ad hoc} way anyhow  (cf.
e.g.  a `static'  prescription for $  M_{\Sigma_c} - M_{\Lambda_c}$
below).

Let us recall some basic features of this model
developed by De Rujula, Georgi and Glashow \cite{Gleshow} long ago.
It is assumed that the hadron consists of 2 (meson) or 3 (baryon)
constituent quarks.
The
quarks in this model have `effective' masses,  e.g. $m_c\sim 1.5$
GeV, $m_u\sim 0.34 $ GeV, etc.  In order not to mix
them
with the current quark masses we  denote them as $m^*$.  The
quarks
are described by nonrelativistic spinors and are bound by a
nonrelativistic
potential modified by hyperfine interactions. The hamiltonian is
equal to
\begin{equation}
H=L({\vec r}_1,{\vec r}_2...)+\Sigma (m^*_i+\frac{\vec
p_i^2}{m^*_i})+\Sigma
(\alpha Q_iQ_j+k
\alpha_s)S_{ij} .
\label{eq:int}
\end{equation}
Here $m^*_i$ are the constituent quark masses, $\vec p_i$ are the
quark
momenta,
and ${\vec r}_i$ are their coordinates; $Q_i$ are the quark charges,
$\alpha_s$
is the QCD coupling constant, and $S_{ij}$ denotes a relativistic
hyperfine
interaction. The constant $k=-4/3$ for mesons and $k=-2/3$ for
baryons.
The potential $L$ is a universal  quark-binding potential.
 The QCD interactions due to the gluon exchange are modeled by the
Breit-Fermi
interaction $S_{ij}$  assumed to have the same form as the
electromagnetic Breit-Fermi  interaction $S_{ij}$
 We refer the reader to the original paper \cite{Gleshow}
for a detailed discussion of this model and  explicit (but rather
complicated ) expressions for $S_{ij}$.
Using the mass dependence of the hyperfine interactions it is
possible to
show that in this model the absolute values of the squares  of the
wave
functions of the baryon and the meson at the origin are connected as
follows:
\cite{Gleshow,Cortes}
\begin{equation}
\frac{\vert \psi (0)^{(cd)}_{\Lambda_c}\vert ^2}{\vert \psi
(0)_D\vert
^2}
=\frac{2m^*_u(M_{\Sigma^+_c}-M_{\Lambda_c^+})}{(M_D-
m^*_u)(M_D-
M_{D^*})} .
\label{eq:newtrick}
\end{equation}
\par Following  refs.
\cite{Voloshin1,Voloshin2,Tadic} let us try to extract as much
information as we can from the nonrelativistic quark
model concerning the 4-quark operators under investigation.
 As we already noted above, the price will be the appearance in our
equations of some parameters which are not connected
in any way with the basic QCD, such as a constituent quark mass.
Consider, for example, the  $\Lambda_c$ baryon.
\par We begin from the matrix elements of the operator
$\bar c \Gamma_\mu c\bar d \Gamma_\mu d$. These matrix
elements are
proportional to $\vert \psi (0)^{(cd)}_{\Lambda_c}\vert ^2$, the
square of the
baryon
wave function at  the origin. More exactly, this is the probability for
$c$ and $d$ quarks to meet. The coordinate of the third quark is
integrated over.  From now on the superscript $cd$ will be omitted.
The  relation
(\ref{eq:newtrick})
connects  the squares of the wave functions in  the
baryons
and mesons \cite{Cortes,Tadic} and will be systematically applied to
make estimates in baryons. Note that in the limit
$m_c\rightarrow\infty$ we are interested
in
$2M_D(M_{D^*}-M_D)=\mu^2_G\times 4/3$ and $M_{\Sigma^+_c}-
M_{\Lambda_c^+}$
tends to a constant value.
Consequently,
\begin{equation}
\vert \psi (0)_{\Lambda_c}\vert^2\sim \frac{3(M_{\Sigma^+_c}-
M_{\Lambda_c^+})
  m^*_u}{\mu^2_G}|\psi (0)_D |^2, \,\,\,
|\psi (0)_D |^2 =\frac{1}{12}F_D^2M_D\kappa^{-4/9} .
\label{eq:baryon}
\end{equation}

Recall now \cite{Voloshin2,Tadic} that the matrix element
of the operator $\bar c \Gamma_\mu c\bar d\Gamma_\mu d$ over
the baryon
$H_c$
is equal to
\begin{equation}
\frac{1}{2M_{H_c}}<H_c\vert  \bar c \Gamma_\mu c\bar
d\Gamma_\mu d\vert H_c>
=\vert \psi_{H_c} (0)\vert
^2 (1-4\vec S_c\vec S_d)
\label{eq:matrix}
\end{equation}
where $\vec S$ denotes the spin operator.
In baryons from  the antitriplet -- $\Xi_c$ and $ \Lambda_c$ --
there is no
correlation
between the spin of the heavy quark and the spins of the light
quarks.
Hence the spin part of eq. (\ref{eq:matrix}) is zero. Using the
expression for
$\vert \psi(0)_D\vert$ in terms of $F_D$ and $M_D$ we immediately
obtain
\begin{equation}
\frac{1}{2M_{\Lambda_c}}<\Lambda_c\vert  \bar c \Gamma_\mu
c\bar d\Gamma_\mu d\vert
\Lambda_c>
\sim \frac{1}{4\mu^2_G}(M_{\Sigma^+_c}-
M_{\Lambda_c^+})m^*_uF^2_DM_D\kappa^{-4/9}.
\label{eq:lambda}
\end{equation}
In the same way we find for the sextet $\Omega_c$ baryon
\begin{equation}
\frac{1}{2M_{\Omega_c}}<\Omega_c \vert \bar c \Gamma_\mu c\bar
d\Gamma_\mu d\vert
\Omega_c >
\sim \frac{5}{6\mu^2_G}(M_{\Sigma^+_c}-
M_{\Lambda_c^+})m^*_uF^2_DM_D\kappa^{-4/9}.
\label{eq:alambida}
\end{equation}
Here we assumed that the spin splittings   like $M_{\Sigma^+_c}-
M_{\Lambda_c^+
}$ do not depend on flavor and are the same for all charmed baryons.

The matrix elements of other operators over $\Lambda_c$ can be
expressed
via matrix element (\ref{eq:lambda}) using the Fierz identities and
antisymmetry
in color of the baryon wave functions.
In particular, using the color antisymmetry of the baryon wave
functions
 we obtain that \cite{Voloshin2,Tadic}
\begin{equation}
<\Lambda_c \vert \bar c_i \Gamma_\mu c^j\bar d_j\Gamma_\mu
d^i\vert \Lambda_c >
=<\Lambda_c \vert \bar c \Gamma_\mu d\bar d\Gamma_\mu
c\vert
\Lambda_c >
=-<\Lambda_c \vert \bar c \Gamma_\mu c\bar d
\Gamma_\mu d\vert \Lambda_c > .
\label{eq:secondpart}
\end{equation}
Similar relations can be found for all other charmed baryons.

 Note that if we want to pursue a systematic expansion
in $1/m_c$ we must use the static value of  the  meson decay
constant
$F_D$ in all expressions above.

\section{Static Versus Actual $f_Q$ and Speculations
on Higher Order Corrections}

\par Above we have carried out  a  systematic expansion of the
inclusive width in  $1/m_Q$ including the contributions of
all
operators with dimensions up to 6.
Let us consider our estimates of dimension 6 four-quark operators
more
carefully.
As it was already mentioned if we want to have a systematic
$1/m_Q$ expansion we must throw away  the part of $f_D$ that
contains
the terms suppressed by powers $1/m_Q$ and use the
static meson decay constant $F_D$ in all our estimates.
If we want to deviate from this procedure we  must take into
account the terms of dimension 7 and higher.

A natural question is whether we have to deviate from this
procedure.
After all, it is reasonable to expect that the contribution suppressed
as $1/m^4_c$ will be unimportant.
Unfortunately,
this is not the case. If we combine  the formulae of section 2.5
with the estimates of the 4-quark operators obtained using $F_D$,
we obtain that they give the contribution to the total width
of $D^+$ meson which is negative and $\sim$ 3 $\times$ the width
 of $D^0$ meson. This is clearly nonsense, meaning that the
higher dimensional operators play an important role in
the hadronic width of $D^+$ mesons. Can we model in any way the
contribution of the higher-dimension operators? Can we at least
argue
what they must show up in order to have sensible results?

 First, consider the mesons.
We shall argue that in the meson case there is at least a set of
higher-dimension operators whose effect
is to renormalize $F_D$ to $f_D$.
\par Indeed, consider the operators  appearing  in the expansion of
the four-quark operators in terms of the operators of
HQET.  Recall that heavy quark currents like
$\bar c \Gamma c$ have the following  expansion :
\begin{eqnarray}
\begin{array}{cl}
(2M_D)^{-1}\bar c \Gamma c=&
\bar h_c \Gamma h_c \\
&+(i/2M_D)({\bar h}_c
 \Gamma \hat Dh_c-\bar h_c
\stackrel{\leftarrow}{\hat D}\Gamma h_c)+O(1/m^2_c).
\label{eq:short}
\end{array}
\end{eqnarray}
Here $\hat D\equiv \gamma^\alpha D_\alpha $.
The four-quark operators  we are
interested in also have a similar expansion,
\begin{eqnarray}
\begin{array}{cl}
\frac{1}{2M_D}\bar c \Gamma_\mu c\bar q\Gamma_\mu q&=
(\bar h_c \Gamma_\mu h_c\bar q\Gamma_\mu q
\\
&+(i/2M_D)(\bar c
 \Gamma_\mu \hat Dh_c-\bar h_c
\stackrel{\leftarrow}{ \hat D}\Gamma h_c))\bar q\Gamma_\mu
q+O(1/m^2_c).
\label{eq:short4}
\end{array}
\end{eqnarray}
 The matrix element of the operator (53) is naturally
expanded in terms of the HQET operators. The matrix element
of
the first term in this expansion is equal to
\begin{equation}
<D\vert\bar h_c \Gamma_\mu h_c\bar q\Gamma_\mu q\vert
D>=F^2_DM_D/6.
\end{equation}
We used the Fierz transformation and factorization to estimate
the latter matrix element.
\par By the same token, assuming that factorization still holds for the
next terms we obtain that the matrix element of the
second term in the expansion (\ref{eq:short4}) is
such that
\begin{equation}
\frac{1}{2M_D}
<D\vert\bar c \Gamma_\mu c\bar q\Gamma_\mu q\vert
D>=\frac{1}{6}
M_DF^2_D(1+2C_1/M_D+...).
\end{equation}
The right hand side of the latter equation is nothing else than the
first two terms in the $1/m_c$ expansion of $f^2_DM_D$.
\par The above arguments show that it is natural to assume that
within factorization the
role of the higher dimensional terms is to renormalize $F_D$ to
$f_D$  in calculating the 4-quark matrix elements. Unfortunately,
the situation is not that  simple. First, there are other terms of
higher
dimensions beyond the ones that appear in the HQET representation
of the four-quark operator. Second, even among the operators
that appear in this representation
 there are "nonfactorizable" ones, that can not be reduced to the form
of
the
product of the light quark current -- heavy quark current. We can
only
hope that
future investigation will clear the situation.
\par Note that the need of  renormalization of $F_D$  directly
follows
from the hope  that our theory  has sense in the charmed family.
Indeed, if we use $F_D$
in the estimates of the 4-quark operators in mesons,
 we obtain that the width of $D^+$ meson is negative ($\sim -2$
widths of the
$D^0$
meson), a clear nonsense. On the other hand, if we use $f_D$,
we obtain  something more reasonable and  in  accord with the
 data.
\par Following the  above discussion
we conjecture that the effect
of the higher dimension terms  is
to renormalize the leptonic decay constant of the heavy hadron in
the
estimate of the 4-quark matrix elements from its static to its real
value provided that factorization is applicable. We use the
`real world'  value of the leptonic decay constant while doing calculations in
mesons.

You can view the speculation above ($f_D$ versus $F_D$) as just an
attempt to disentangle the overall normalization of the mesonic
matrix elements from that of the baryonic matrix elements, a
connection inherent to NQM, if
you are  not inclined to take it more seriously.
\par Consider now the case of the baryons. In this case we can not
repeat
the above calculations -- for baryons  factorization is not applicable.
We can not estimate the
contribution of
the operators appearing in the second line of  (53). We,
thus,
have to be satisfied with three first orders in $1/m_Q$ expansion,
and
have no alternative but to use the static constant
$F_D$ in
our calculations {\em assuming} that here the higher order terms
do not play as drastic role as in the meson lifetimes.  Quite
remarkably, we will see later on that
then we do get  the experimental pattern of the lifetimes.

\par We end this section by reiterating that the preasymptotic power
terms must be much larger in baryon lifetimes than in meson one,
and this can be achieved by
 using   $F_D$ for baryons and $f_D$
for
mesons.  The difference is formally of higher order in $1/m_c$  -- it
is necessary to have enhanced terms ${\cal O}(m_c^{-4})$ in mesons
and suppressed in baryons.

\par An obvious difference between the baryon antitriplet and
mesons
is the presence of the strong spin terms in the latter. The same
strong spin terms are present in the $\Omega_c$ baryon. It is,
perhaps, possible to conjecture that an unknown mechanism that
distinguishes
between the mesons and the antitriplet baryons is somehow
connected with
the spin terms. If so, a  natural consequence might be that the
charmed
heparin $\Omega_c$ has the lifetime  around that of $D^0$, i.e.
larger than the lifetimes of $\Lambda_c$
or $\Xi^0_c$. It will be very interesting to determine this lifetime
experimentally.

\section{Numerical Analysis in the Charm Family}

\par We move now to numericals. We shall consider first the case of
the charm
family. The following numerical values of different parameters are
accepted.
 We use the values $c_+\sim 0.74,\quad c_-\sim 1.82$
corresponding to the currently accepted value of $\Lambda_{\rm
QCD}\sim 300 $. We use $f_D\sim 170$ MeV for the `real world'
value of $f_D$, and $F_D\sim 400$ MeV for
the static leptonic decay constant, both normalized at $m_c$
\cite{F,F1}.
 We  choose the  normalization point $\mu\sim
 0.5$
GeV for the estimates of matrix elements of the four-quark
operators.
The
corresponding value for $\kappa$ is $\kappa\sim 3.1$.
We use the numerical values $m_c\sim 1.4$ GeV, $\mu^2_G\sim
0.4$ GeV$^2$,
and $\mu^2_\pi\sim 0.5$ GeV$^2$.

We also use $M_{\Sigma^+_c}-M_{\Lambda_c}=400$ MeV,
not the real world value $M_{\Sigma^+_c}-M_{\Lambda_c}=200$
MeV,
in the systematic $1/m_c$ expansion,
since, as it
follows from QCD sum rules \cite{Shuryak}, this mass difference goes
to
$\sim $ 400 MeV  in the limit of the infinite mass of the heavy
quark. (Remember, this is just an overall normalization of the NQM
matrix elements, see eq. (46)).
 For the constituent quark mass we use  $m^*_u\sim 0.35 $ GeV.
\par In the naive parton model (the dimension 3 operator ${\cal
O}_0$) all hadrons
have the same width
\begin{equation}
\Gamma =N_c\eta\Gamma_0\sim 4.5\Gamma_0.
\label{eq:parton}
\end{equation}
They also have the same semileptonic widths
\begin{equation}
\Gamma_{\rm s/l}\sim 2\Gamma_0.
\label{eq:s/l}
\end{equation}
The total width in this model is $\sim 6.5\Gamma_0$.

Consider next the contribution of dimension 5 operators.
We begin from the charmed mesons. These operators are spectator
blind and
contribute in the same way to $D^0$ and $D^+$.  Using the explicit
formulae
for the matrix elements of ${\cal O}_0, {\cal O}_G, {\cal O}_\pi$
and the Wilson coefficients $C_0, C_G, C_\pi$
it is easy to obtain that the dimension 5 terms from
${\cal O}_0$, ${\cal O}_G$ and $ {\cal O}_\pi$ shift
hadronic widths of D mesons by
\begin{equation}
\Delta \Gamma^{(5)}\sim 2.5 \Gamma_0.
\end{equation}
Let us now consider in more detail the case of $D^0$ meson.
Recall that its width also gets contribution from annihilation
mechanism $\sim 0.9\Gamma_0$ \cite{BlokShifman}, and that its
semileptonic
width is suppressed by a factor $\sim 1/2$ \cite{Bigi,DPF}.
Then the hadronic width of $D^0$ meson is
\begin{equation}
\Gamma_{\rm hadr}(D^0)\sim 8 \Gamma_0.
\end{equation}
and its total width is
\begin{equation}
\Gamma (D^0)\sim 9 \Gamma_0.
\end{equation}
 Of course, all these estimates have only qualitative  character
although we
certainly are always glad when we see the right tendency. To give a
feeling of the uncertainty, note that if we use $m_c\sim
1.35$
GeV, the width of $D^0$ increases by $\sim 10\%$.
\par This is all with the $D^0$ meson, since the
4-quark
operators of dimension 6 do not contribute to its width.

Let us parenthetically note that $D_s$ is in the same situation as
$D^0$ and is predicted to have the same width in the limit when we
neglect
$SU(3)$ breaking effects. A tiny difference between $D_s$ and $D^0$
is exhaustively discussed in a recent work \cite{UB1}, and we  will
not dwell on this point here.

\par For the baryons (except $\Omega_c$) the contribution of
dimension 5
operators
is due only to ${\cal O}_\pi$ operator and is $\sim -0.5\Gamma_0$.
The chromomagnetic operator
$O_G$ proportional to the correlations between
the
spin of the heavy quark and the spin of the light diquark does not
show up since the spin of the latter
is
zero for the baryon antitriplet. For $\Omega_c$ we estimated
the contribution of this operator in section 4.2.

We now move to dimension 6 operators.
 The values of the Wilson coefficients
$K_i\eta_i$, eq.
(\ref{eq:factors}), are:
\begin{eqnarray}
\begin{array}{cl}
K_1\eta_1
&\sim 2.28,\\
K_2\eta_2&\sim -2.45,\\
K_3\eta_3&\sim -0.88,\\
K_4\eta_4&\sim 0.45,
\\
K_5\eta_5&\sim 0.31,\\
K_6 \eta_6&\sim
-2.65.
\label{eq:factorsn}
\end{array}
\end{eqnarray}

Consider first the charmed mesons.
 The contribution of the 4-quark operators in the width of  $D^0$
meson
is zero. For $D^+$ meson the contribution of the 4-quark operators is
given by
\begin{equation}
\Gamma^{(6)}=48\pi^2\Gamma_0(\frac{K_1\eta_1}{3}+K_2\eta_2)
\frac{f_D^2}{M^2_D}
\frac{1}{\kappa^{4/9}}.
\label{eq:old}
\end{equation}
Here $\kappa^{-4/9}$ takes into account the hybrid
logarithms
that appear in $f_D$ when
we  move
between $m_c$ and a low  normalization point.
Substituting the numerical values of the parameters given above and
using the "real world" value of the  constant $f_D\sim 170$
MeV
we obtain that the
contribution of the dimension 6 operators into the width
of the $D^+$ meson is
\begin{equation}
 \Delta\Gamma^{(6)}(D^+)\sim -4\Gamma_0.
\end{equation}
For the total and hadronic widths of the $D^+$ meson we immediately
obtain (taking into account the suppression of the semileptonic width
mentioned above, and the contributions of ${\cal O}_0$ and the
dimension 5 operators that are the same as for $D^0$ meson)
\begin{equation}
\Gamma_{\rm hadr}(D^+)\sim
4\Gamma_0 ,\quad
\Gamma (D^+)\sim 5\Gamma_0,
\end{equation}
\begin{equation}
\frac{\Gamma (D^0)}{\Gamma (D^+)}\sim 2.
\end{equation}
This result in an agreement with the experimental pattern, as can be
easily seen from  Table 1.
\par This agreement, however, must not be taken too literally.
Indeed, in
the latter estimate we neglected the difference between $M_D$ and
$m_c$
in the denominator of  eq. (\ref{eq:old}). Were we distinguishing
between $M_D$ and $m_c$  we would obtain, instead of
the factor
 $f^2_D/M^2_D$ in eq. (\ref{eq:old}),
the factor $f^2_DM_D/m^3_c$ that will give a huge contribution
of dimension 6 operators, close to  $-7\Gamma_0$, reducing
the width
of $D^+$ almost to zero. This once again shows the importance of the
contribution of higher-dimension corrections. In our case it seems
logical to substitute $m_c$ by  $M_D$ in the estimates since we use
the real
world value of $f_D$.
If we used the systematic $1/m_c$
expansion and the static value of leptonic decay constant  $F_D$
we would obtain an unphysical negative width of $D^+$ meson. Alas,
the charmed hadrons are too light!
\par Let us now consider the baryons.
Using the explicit expressions for the baryon matrix elements in the
NQM
it is straightforward to get the following expressions for the
matrix elements of the dimension 6 operators:
\begin{eqnarray}
\begin{array}{cl}
W_d&\equiv \frac{1}{M_{H_c}}<H_c\vert{\cal L}_d\vert
H_c>=(K_1\eta_1-
K_2\eta_2)\delta\Gamma_B,\\
W_u&\equiv \frac{1}{M_{H_c}}<H_c\vert {\cal L}_u\vert
H_c>=(K_3\eta_3-K_4\eta_4)
\delta\Gamma_B,\\
W_s&\equiv
 \frac{1}{M_{H_c}}<H_c\vert{\cal L}_s\vert H_c>=(K_5\eta_5-
K_6\eta_6)\delta\Gamma_B.
\label{eq:contr}
\end{array}
\end{eqnarray}
 Here we neglected the penguin-type
structures.
The width $\delta\Gamma_B$ is determined by the following
expression
\begin{equation}
\delta\Gamma_B=48\pi^2\Gamma_0\vert \psi^2(0)\vert
2m_c^{-3}\sim
48\pi^2\Gamma_0 \frac{m^*_uF^2_DM_D(M_{\Sigma^+_c}-
M_{\Lambda_c})}{2\mu^2_Gm^3_c\kappa^{4/9}}.
\label{eq:4}
\end{equation}
The value of $\delta\Gamma_B$ depends on whether we use $F_D$
and
what value of
$(M_{\Sigma^+_c}-
M_{\Lambda_c})$ is substituted. If the static mass difference,
$400 $ MeV, is
substituted, along with the static $F_D$,
we immediately obtain  that the contributions of the different 4-quark
operators are:
\begin{eqnarray}
\begin{array}{cl}
W_d&\sim 28\Gamma_0,\\
W_u&=\sim -8\Gamma_0,\\
W_s&\sim 22\Gamma_0.
\label{eq:cons}
\end{array}
\end{eqnarray}
For the contributions of the 4-quark operators into the hadronic
widths of the
charmed baryons we find
\begin{eqnarray}
\begin{array}{cl}
\Gamma^{(6)}_{\rm hadr}(\Lambda_c^+)&= W_d+W_u\sim
20\Gamma_0,\\
\Gamma^{(6)}_{\rm hadr}(\Xi^+_c)&=W_u+W_s\sim  14\Gamma_0 ,
\\
\Gamma^{(6)}_{\rm hadr} (\Xi^0_c)&=W_d+W_s\sim 50\Gamma_0 .
\label{eq:4widths}
\end{array}
\end{eqnarray}
Let us now consider the full baryonic and mesonic widths. Taking
into
account that the  semileptonic  widths of
the triplet baryons  are approximately the same as in the naive
parton
model (no ${\cal O}_G$ contribution)
and assembling together  operators ${\cal O}_0$, ${\cal O}_\pi$
and 4-quark contributions
we immediately obtain
\begin{eqnarray}
\begin{array}{cl}
\Gamma(\Lambda_c^+)&= 26\Gamma_0\sim 5\Gamma (D^+),\\
\Gamma(\Xi^+_c)&= 20\Gamma_0\sim 4\Gamma (D^+)\sim
2\Gamma (D^0) ,\nonumber\\
\Gamma (\Xi^0_c)&=56\Gamma_0\sim 11\Gamma (D^+).
\label{eq:5widths}
\end{array}
\end{eqnarray}
These results are in a qualitative
 agreement with the experimental pattern in Table
1.
\par
 If instead of the static $F_D$  we used $f_D$, we would get numbers
completely contradicting  the experimental pattern  and leading  to
too narrow a span of the lifetimes.
\par We conclude that the scale of the matrix elements of the
four-quark operators  must be enhanced in baryons compared to
mesons.
Within NQM we achieve this enhancement by applying a prescription
$f_D\rightarrow F_D$.
{}From theoretical point of view the obvious difference between the
mesons and the
antitriplet
baryons is the absence of the interactions between heavy quark spin
and that of the light cloud
 in the latter. Is this the underlying reason for the
different
behavior of the heavy quark expansion in these two cases? There
exists
one charmed baryon, namely $\Omega_c$, that has a strong
spin-spin
interaction like in the mesons.
If we use the "static" prescription, as for antitriplet baryons, we
obtain that the width of this baryon is $\sim 16 \Gamma (D^+)$.
If we follow the meson example and use the $f_D$ prescription,
we obtain that its width is $\sim 2\Gamma (D^+)$. Depending on the
prescription used $\Omega_c$ can be both the most short-living and
the most long-living
among charmed baryons.
 There is yet no experimental
data on the width of this baryon. Such data can be crucial to
establishing the QCD mechanism of the hierarchy formation.

Note the main distinctions between our analysis and the analysis
carried in
refs. \cite{Voloshin1,Voloshin2,Tadic}. In comparison with
the first two works, we used different values of the Wilson
coefficients.
The ones used in \cite{Voloshin2} correspond to
$\Lambda_{\rm QCD}\sim
100$ MeV. New values of the Wilson coefficients change in
the
right way the relative contributions of different four-quark
mechanisms
(proportional to $W_d,W_u,W_s$, respectively) in the
inclusive
widths of the baryons. We also discarded a rule of discarding
$1/N_c$ accepted in \cite{Voloshin2}.  Instead,  we took into account
the contribution of dimension
5
operators. Next, following the authors of ref. \cite{Tadic} we used
the relation \ref{eq:newtrick} between the absolute values of the
wave
functions
of
baryons and mesons. Finally, we used the static value $F_D$
for
the estimates of the baryonic matrix elements.
\par In comparison with the analysis \cite{Tadic} we included the
hybrid logarithms and
the contribution of dimension 5 operators.

\section{Estimates for Beautiful Hadrons}
 The analysis for beautiful hadrons proceeds in exactly the same
way as
for charmed ones.
Moreover, we expect our predictions to be more accurate, since the
expansion parameter is $\sim 0.1$.

The only difference with the charm case is the presence of the phase
space factors:
$f_0=1-8r+8r^3-r^4-12r^2\ln r$ for the $\bar cc$ operator and
$(1-r)^3$ and $(1-r)^4$ for ${\cal O}_G$ and ${\cal O}_\pi$ operators
respectively.
We use the $b$-quark mass $m_b\sim 4.8$ GeV and the values
$f_B\sim 150$
MeV, $F_B\sim 250$ MeV \cite{F,F1}.
The normalization point $\mu\sim 0.5$ GeV,
corresponding to
$\kappa\sim 5.42$ is chosen. The  values of the Wilson
coefficients are
$$c_+=0.84, \,\,\, c_-=1.41.$$
\par The question of the lifetime hierarchy is especially interesting,
since
in this case we expect the inverse heavy quark mass expansion to
work much
better than for  the charmed hadrons. All calculations become
more reliable. It will be of great interest to see whether the approach
becomes quantitative for the beautiful hadrons. Recall that we
speculated that  $f_D$ is more relevant for mesons while  $F_D$ is
more relevant for baryons.  The difference between $f_B^2$ and
$F_B^2$ is still quite noticeable -- about 40\%. The same
regularity is expected to  show up in the  beautiful  hadrons.
Unfortunately, at this point we
do not
have any reliable experimental data, so we just consider theoretical
predictions that follow from the use of the same rule for beauty as
for
charm.
\par Using the values of the parameters given in the beginning of
this
section,
we immediately obtain rather large lifetime differences between
different
beauty mesons. In particular,
 we obtain that the lifetimes of charged and neutral $B$
mesons
can differ by as much as  3 to 5\%. The width of
the charged meson $\Gamma(B^-)$ decreases
due to the interference effects.
 On the other hand
we obtain that the width of $\Lambda_b$ can be as much as 10$\%$
bigger than the
width of
the B$^0$-meson and is approximately equal to the width
of the $\Xi^+_b$ baryon.
The width of $\Xi^0_b$ baryon is $\sim 3\%$ smaller than the width
of $B^0$
meson.
It will be extremely interesting to check these
predictions experimentally.

     Note that the significant difference with the previous estimate of
the beauty
lifetimes \cite{Voloshin2} comes because we use the value of $f_B$
 $\sim 1.5$ larger  than
Voloshin and Shifman, as it follows from the recent QCD sum rule
analysis
\cite{F}.

\section{Conclusions}
In this talk we addressed the question whether it is possible to
`explain'
the observed hierarchy of the hadronic widths in the charmed
system in
QCD on the basis of the
 heavy quark expansion. Strictly speaking the charm family seems
to lie below or, with luck, at the border of the domain where the
expansion may be useful,
which makes the task absolutely challenging. We speculated that
there is a
significant
difference in the structure of the preasymptotic expansion in the
meson and baryon systems. Terms of higher dimensions show up
in a completely different way in mesons and baryons. This situation
can be effectively modeled by taking into account the first  orders
of the
preasymptotic expansion (up to ${\cal O}(m_c^{-3})$) and by
 using the real world value of the  constant $f_D$ in mesons and the
static $F_D$ in baryons. If such a picture  really exists  the problem
of a wide spread of the total  widths in the charmed family is solved,
 at least at the qualitative level.
\par Unfortunately, the origin of this basic difference between
mesons and baryons
is still unclear. Since the obvious difference between mesons and
the
antitriplet baryons is the absence of the spin-spin interaction in the
latter (and its presence in the
former), it will be extremely important to get a reliable data on
the lifetime of the sextet baryon $\Omega_c$. This is the only baryon
where we expect that the spin-spin interactions play a major role, as
in
mesons.
\par Assuming that the same mechanism works also for beauty
hadrons, we
listed in the last section some  estimates for beauty hadrons.
 It is extremely interesting to check them in
current
experiments.

Finally we must stress that our estimates are just provocative.
 Our task here was , first of all, to formulate
the problem of the hierarchy, to see what is known in the current
literature about it, and whether it can be solved using  available
methods. We have speculated how  QCD must work in order to
explain the
hierarchy. The next question is  whether it really works that way.

{\bf Acknowledgments}
It is a pleasure to thank I. Bigi for interesting and stimulating
discussions. This work is supported in part by DOE under the contract
DOE-AC02-83ER40105.

\newpage

\begin{center}

{\bf Table 1}
\vspace{0.25cm}

{\bf Experimental Data on Weak decays of Charmed Hadrons}
\vspace{0.25cm}

\end{center}
\begin{tabular}{|l|l|} \hline
\multicolumn{1}{|c|}{~~{\bf hadron}} &
\multicolumn{1}{c|}{~~{\bf lifetime}~~~} \\ \hline
& \\
$D^+$ & $\sim 10.6\times 10^{-13}$ s  \\
 & \\ \hline
& \\
$D^0$ & $\sim  4.2 \times 10^{-13}$ s \\
& \\ \hline
& \\
$\Lambda_c$ & $\sim 2\times 10^{-13}$ s \\
& \\ \hline
& \\
$\Xi^+_c$ & $\sim 3\times 10^{-13}$ s \\
& \\ \hline
& \\
$\Xi^0_c$ & $\sim 0.8\times 10^{-13}$ s \\
& \\ \hline
\end{tabular}
%\eject
\newpage
\centerline{\bf Figure Captions.}

{\bf Fig. 1:} The  imaginary part of this graph determines the hadronic
width of all $c$ containing hadrons in the parton model approximation. The
closed circle denotes the effective weak lagrangian.

{\bf Fig. 2:} The influence of the soft modes. A soft gluon (absorbed by
the light cloud of $D$) is marked by x.

{\bf Fig. 3 a,b,c:} The diagrams giving rise to ${\cal O}_{4q}$ in
$\hat T$. The crosses mark soft $d$, $u$ or $s$ quark lines

\end{document}